\documentclass{optica-article}

\journal{opticajournal} 

\articletype{Research Article}

\usepackage{lineno}

\begin{document}

\title{Broadband High-Speed Dual-Comb Spectroscopy Enabled by a Dynamic 1550 nm Bidirectional Dissipative Soliton Fiber Laser}

\author{mingjun wang,\authormark{1} zhangru shi,\authormark{1} anshuang wang,\authormark{1} and bowen li\authormark{1,*}}

\address{\authormark{1}Key Laboratory of Optical Fiber Sensing and Communications (Education Ministry of China), University of Electronic Science and Technology of China; Chengdu, 611731, China\\

\email{\authormark{*}bowen.li@uestc.edu.cn} 
}

\begin{abstract*} 
We report a high-energy, bidirectional, dissipative soliton mode-locked fiber laser operating in the 1550 nm normal-dispersion regime. By leveraging intracavity dispersion management and a Lyot filtering mechanism, the laser achieves flat-top optical spectra with 10-dB bandwidths exceeding 20 nm in both directions. Single-pulse energies of 2.7 nJ and 1.5 nJ are achieved for the clockwise and counter-clockwise directions, respectively. Furthermore, the all-fiber configuration exhibits superior noise performance and inherent common-mode noise suppression. To facilitate broadband and high-speed dual-comb spectroscopy, we employ a dynamic repetition rate difference control technique via pump power modulation, enabling zero-crossing dynamic scanning. This approach achieves a spectral measurement bandwidth of approximately 16 nm at an acquisition rate of 500 Hz. Compared to static operation, this represents a nearly two-order-of-magnitude improvement in acquisition speed and achieves a fivefold measurement bandwidth beyond the Nyquist aliasing limit. Experimental results demonstrate that the system maintains robust coherence even under dynamic modulation. By implementing a phase-correction algorithm, a mutual coherence time of 0.5 s is successfully achieved, yielding a spectral resolution exceeding 7.2 GHz. This work fills a gap in high-energy dissipative soliton dual-comb sources at 1550 nm and provides an ideal solution for low-cost, high-sensitivity dual-comb spectroscopy requiring both broad bandwidth and high speed. 

\end{abstract*}

\section{Introduction}
Dual-comb spectroscopy (DCS) circumvents the inherent limitations of mechanical interferometers, offering a robust paradigm for high-speed, high-resolution spectral analysis \cite{diddams2020optical,coddington2016dual}. Recently, the escalating demand for field-deployable and real-time applications has intensified the pursuit of dual-comb sources that simultaneously feature high coherence, low complexity, broad bandwidth, and rapid acquisition capabilities \cite{picque2019frequency}.

Conventional dual-comb systems typically rely on two independent mode-locked lasers coupled with sophisticated phase-locking and synchronization controls. This configuration leads to excessive bulk, high costs, and poor environmental adaptability. In contrast, single-cavity dual-comb fiber lasers (SCDCLs) generate two pulse trains within a single resonator, offering superior compactness and integrability. More importantly, since the two combs share identical noise channels, such as temperature fluctuations and mechanical vibrations within the same cavity, they exhibit outstanding common-mode noise rejection and enhanced mutual coherence \cite{ideguchi2016kerr}. This physical foundation enables dual-comb measurements under free-running or weakly-locked conditions \cite{liao2020dual}, significantly reducing system footprint and cost, and facilitating the transition of DCS from laboratory environments to field applications.

Among the various implementation pathways for SCDCLs, bidirectional mode-locking is particularly advantageous due to its structural simplicity and the provision of two natural outputs. Since the pulses in both directions share the same physical cavity length, they inherently exhibit an extremely small repetition rate difference, which significantly extends the alias-free spectral range, making it an ideal candidate for broadband dual-comb spectroscopy \cite{zeng2013bidirectional,olson2018bi,li2020bidirectional}. Specifically, the dissipative soliton (DS) mechanism in net-normal dispersion fiber lasers is widely regarded as a pivotal route to overcoming the energy limitations of conventional solitons, offering higher intracavity energy and strongly chirped pulses that are externally compressible \cite{grelu2012dissipative,chong2007all}. Leveraging the high pulse energy of DSs, the compressed output can be directly coupled into a nonlinear fiber to generate a supercontinuum, achieving broad spectral coverage without external amplification. This "oscillator-direct-driven" architecture not only simplifies the system link but also ensures high signal-to-noise ratio (SNR) and mutual coherence \cite{camenzind2025ultra}. While high-energy bidirectional DS mode-locked lasers at 1 $\mu$m have been extensively reported [8], achieving similar performance in the 1.5 $\mu$m telecommunication band, which is more critical for molecular fingerprinting \cite{rieker2014frequency}, remains challenging. The anomalous dispersion of standard single-mode fibers in this region forces most existing single-cavity systems to operate in the conventional soliton regime, resulting in limited pulse energy \cite{nakajima2019all,chernysheva2016isolator,zhao2018polarization}. Consequently, constructing a high-energy, broadband, bidirectional mode-locked laser at 1550 nm represents both a fundamental challenge in laser dynamics and a crucial step toward field-deployable, high-sensitivity DCS applications.

We report, to the best of our knowledge, the first asynchronous bidirectional, normal-dispersion dissipative soliton mode-locked fiber laser operating at 1550 nm. The laser delivers DS spectra with a 10-dB bandwidth exceeding 20 nm in both directions simultaneously. At a total pump power of 1.2 W, the single-pulse energies reach 2.7 nJ (clockwise) and 1.5 nJ (counter-clockwise). Beyond the high energy and broad bandwidth typical of DSs, the repetition rate difference ($\Delta f_{\text{rep}}$) can be flexibly and continuously tuned across the zero-point via simple pump power modulation. Compared to traditional schemes requiring complex piezoelectric transducer (PZT) cavity-length control \cite{schliesser2005frequency,kim2010high,shi2022high}, this mechanical-free approach, coupled with the inherent common-mode noise suppression of the single-cavity architecture, enables stable operation without requiring separate repetition rate locking loops for each comb, significantly reducing system complexity. To demonstrate its utility in broadband and high-speed DCS, we implemented a dynamic $\Delta f_{\text{rep}}$ modulation scheme, effectively overcoming the acquisition speed bottleneck typically encountered in broadband measurements. We achieved a spectral bandwidth of $\sim $16 nm at a high acquisition rate of 500 Hz (contrasting a theoretical static alias-free bandwidth of $\sim $3.2 nm). Assisted by a phase-correction algorithm, a mutual coherence time exceeding 0.5 s and a spectral resolution better than 7.2 GHz were attained. This novel 1550 nm DS single-cavity dual-comb source not only fills a critical wavelength gap but also provides an ideal hardware platform for high-performance, cost-effective DCS.

\section{Experimental setup and results}
\begin{figure}[t]
\centering\includegraphics[width=13.5cm]{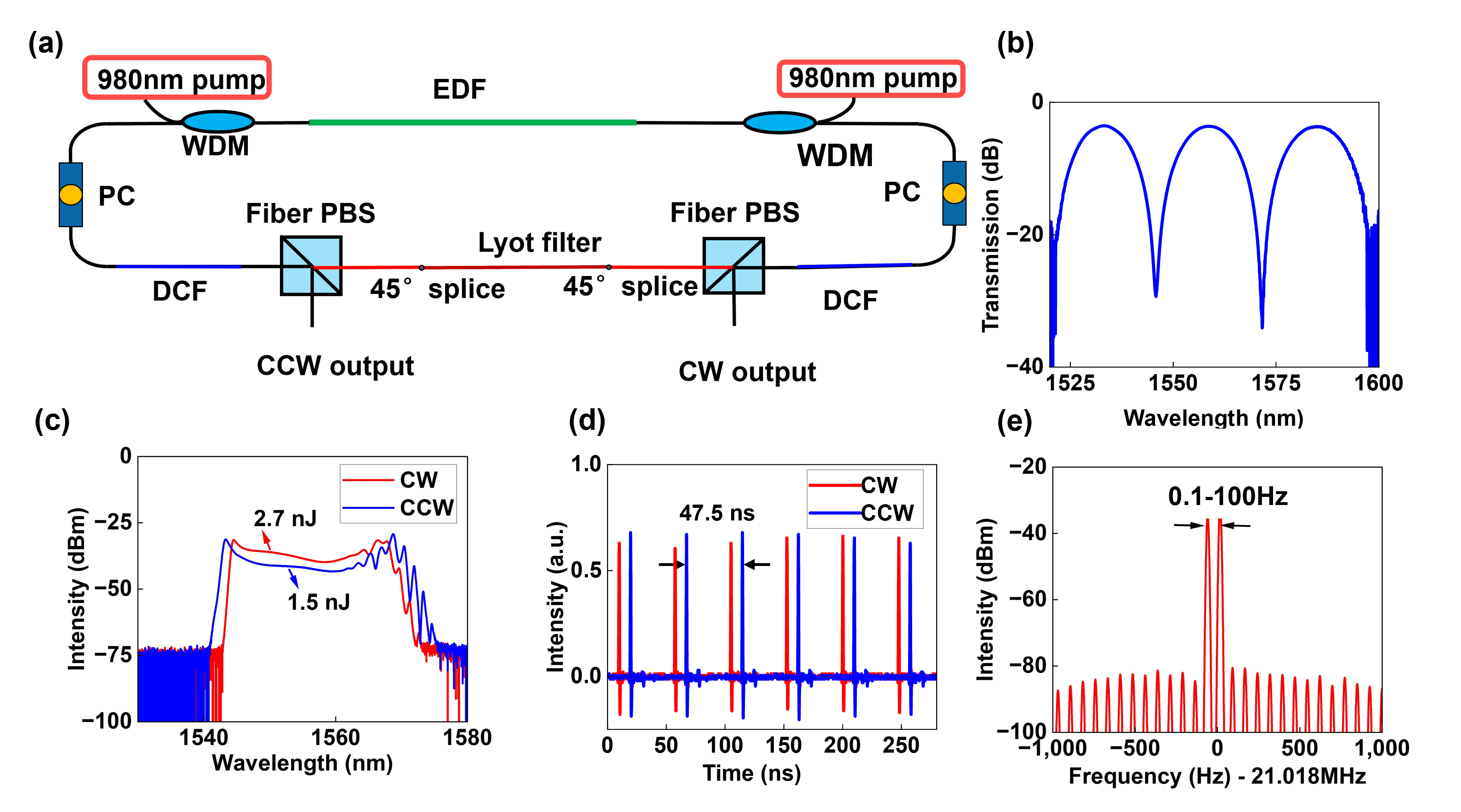}
\caption{(a) Experimental setup of the 1550 nm high-energy bidirectional all-fiber mode-locked laser. (b) Transmission spectrum of the intracavity Lyot filter. (c) Optical spectra of the bidirectional mode-locking state. (d) Pulse trains in the time domain under bidirectional mode-locking. (e) RF spectra of the bidirectional laser output. WDM, wavelength division multiplexer; EDF, erbium-doped fiber; PC, polarization controller; DCF, dispersion-compensating fiber; PBS, polarization beam splitter; CW, clockwise; CCW, counterclockwise;}
\label{fig:1}
\end{figure}

Figure ~\ref{fig:1}(a) illustrates the experimental setup of the 1550 nm bidirectional single-cavity dual-comb fiber laser. To achieve high-energy DS mode-locking in the 1.5 µm band, intracavity dispersion management is employed to counteract the anomalous dispersion inherent in standard fibers. The laser features a ring cavity configuration with a 5-meter L-band normal-dispersion erbium-doped fiber (EDF) as the gain medium, which is symmetrically pumped in both directions by two 980 nm laser diodes via wavelength division multiplexers (WDMs). To establish a net-normal dispersion environment supportive of DS formation, 3.7 meters of dispersion-compensating fiber (DCF) are symmetrically integrated, resulting in a total cavity dispersion of approximately 0.5 ps² and a total cavity length of 9.9 m.

The mode-locking and generation of bidirectional DSs rely on a synergistic mechanism of nonlinear polarization rotation (NPR) and an all-fiber Lyot filter \cite{lyot1944filtre}. The Lyot filter is composed of polarization-maintaining (PM) fibers and a polarization beam splitter (PBS), where the red section represents the PM1550 fiber, spliced at 45-degree angles from three segments. This filter performs sinusoidal periodic filtering, with the modulation period adjustable by varying the length of the middle PM fiber segment \cite{han2022flexible}. As shown in Fig.~\ref{fig:1}(b), with a $\sim $15 cm birefringent PM fiber segment, the filter exhibits a transmission spectrum with a modulation period of approximately 26 nm.

Under this configuration, the laser successfully achieves stable bidirectional DS mode-locking, with typical output characteristics shown in Fig.~\ref{fig:1}(c). Benefiting from the combined effects of strong intracavity normal dispersion and Lyot filtering, both the CW and CCW directions exhibit the spectral characteristics of DSs. The two output spectra are highly overlapped, with 3-dB bandwidths exceeding 20 nm and center wavelengths near 1560 nm. At a balanced pump power of 600 mW for each direction, the single-pulse energies for the CW and CCW directions reach 2.7 nJ and 1.5 nJ, respectively, representing the highest energy levels reported to date for single-cavity dual-comb systems in the 1550 nm band. The temporal pulse trains and the combined RF spectra for both directions are characterized in Figs.~\ref{fig:1}(d) and \ref{fig:1}(e). The pulse sequences recorded by the oscilloscope show a pulse interval of 47.5 ns, corresponding to a fundamental repetition rate of 21.018 MHz. The RF spectra reveal a measured $\Delta f_{\text{rep}}$ of 76 Hz between the CW and CCW directions. Furthermore, by fine-tuning the polarization controller or pump power, $\Delta f_{\text{rep}}$ can be continuously adjusted from -100 Hz to 100 Hz while maintaining uninterrupted mode-locking.

\begin{figure}[t]
\centering\includegraphics[width=13.5cm]{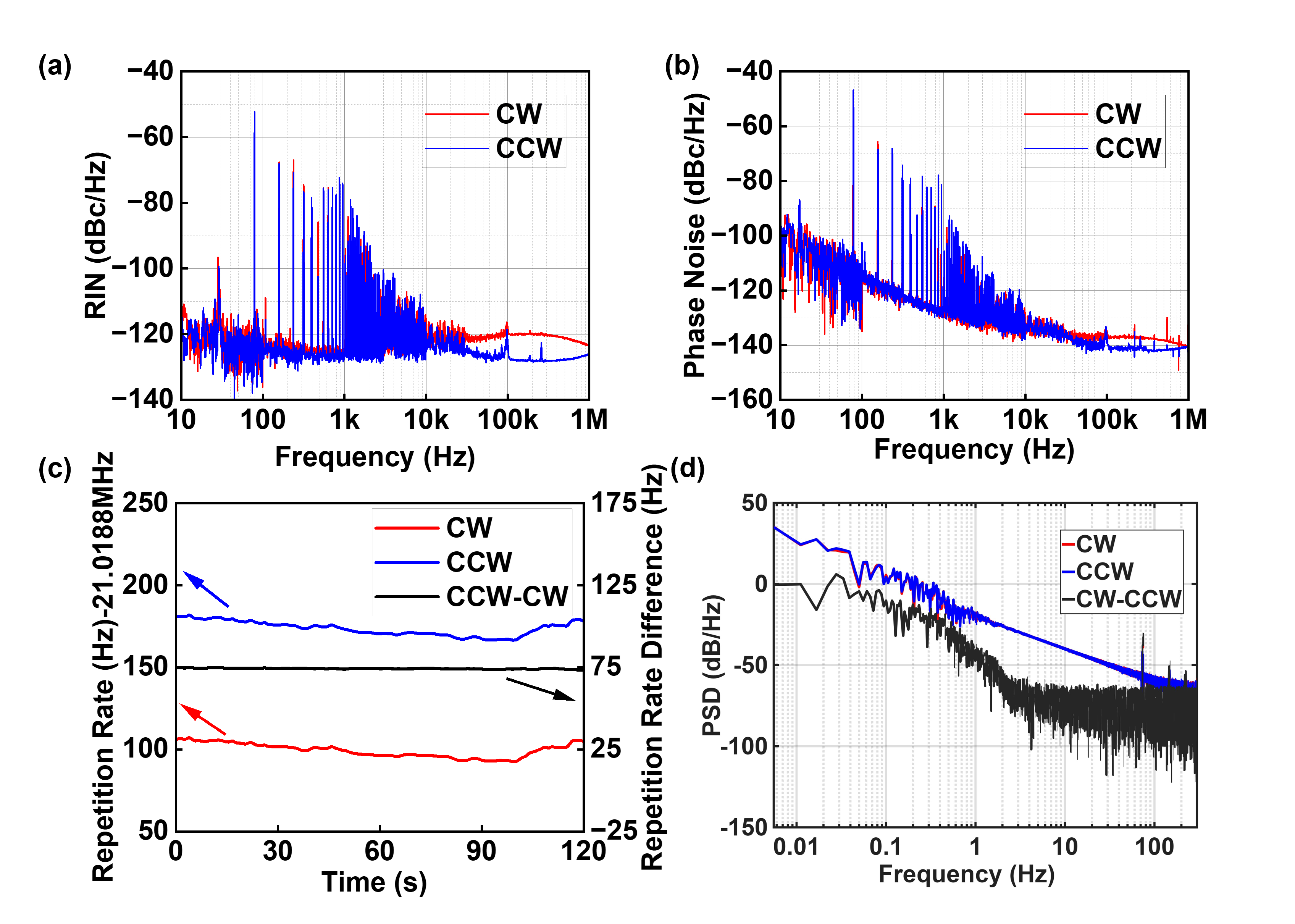}
\caption{Static noise characterization of the 1550 nm dual-comb laser. (a) Relative intensity noise (RIN) and (b) phase noise spectra under bidirectional mode-locking. (c) Long-term stability of the repetition rates. (d) Frequency noise PSD for the repetition rates in the clockwise (CW, red) and counter-clockwise (CCW, blue) directions, and their difference.}
\label{fig:2}
\end{figure}

Figures~\ref{fig:2}(a) and \ref{fig:2}(b) depict the relative intensity noise (RIN) and phase noise spectra for the CW and CCW directions, respectively. Distinct noise peaks are observed at the $\Delta f_{\text{rep}}$ and its higher harmonics in both spectra. Similar phenomena reported in 1 $\mu$m systems suggest that these peaks most likely originate from gain competition between the two combs \cite{li2025comprehensive}. Figure~\ref{fig:2}(c) records the evolution of the repetition rates over a 120-second interval. Although the repetition rates in each independent direction exhibit significant free-running drift, their difference ($\Delta f_{\text{rep}}$) remains remarkably stable. Quantitative analysis reveals that the standard deviation (STD) of the unidirectional repetition rate is approximately 5.26 Hz, whereas the STD of $\Delta f_{\text{rep}}$ is only 0.53 Hz. This order-of-magnitude difference provides compelling evidence for the superior common-mode noise rejection (CMNR) capability of the single-cavity architecture in suppressing environmental perturbations. To further analyze frequency stability in the frequency domain, the power spectral density (PSD) of the repetition rates and their difference is presented in Fig.~\ref{fig:2}(d). The results show a clear CMNR effect of approximately 20 dB in the low-frequency regime (< 10 Hz). Beyond 100 Hz, the measured frequency noise is predominantly limited by the noise floor of the frequency counter.

\begin{figure}[t]
\centering\includegraphics[width=13.5cm]{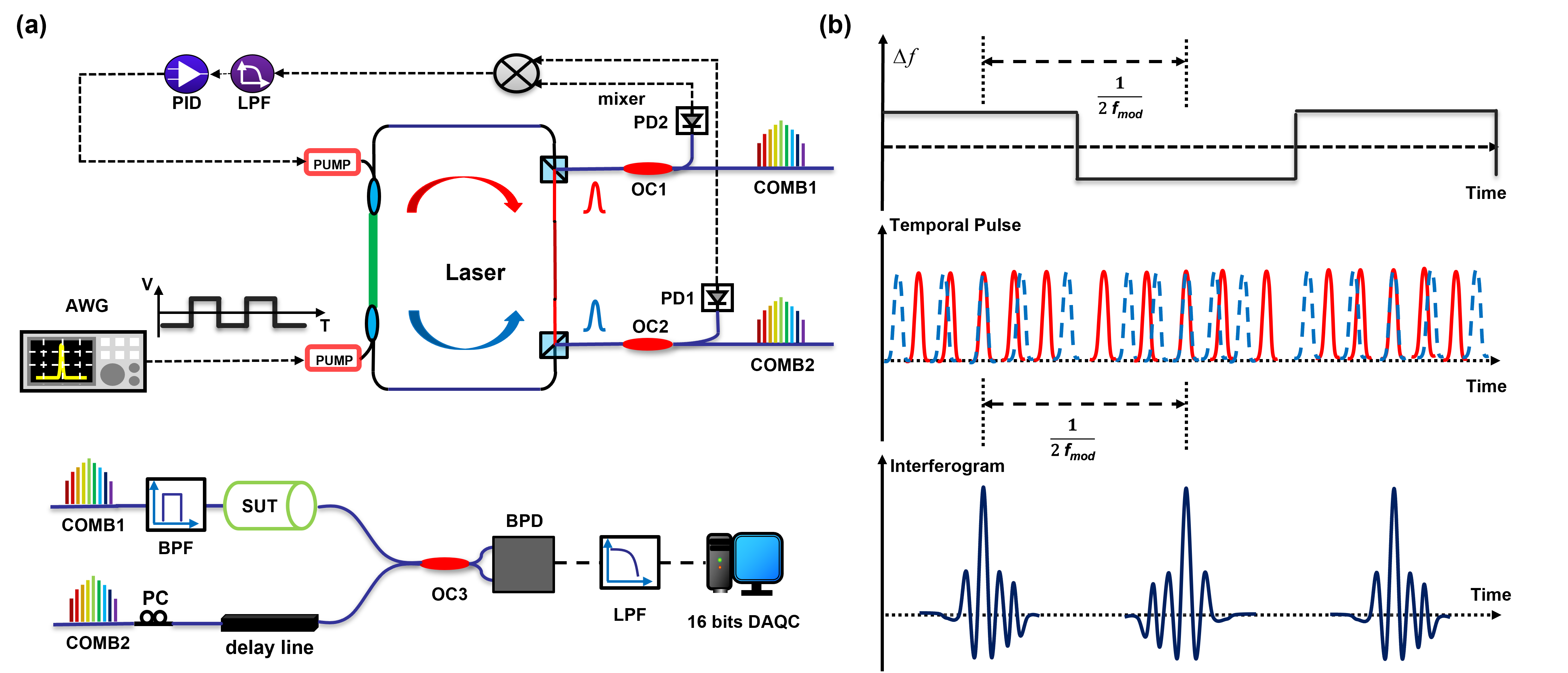}
\caption{(a) Experimental setup for the DCS system with dynamic repetition rate difference control. (b) Schematic concept of the dynamic repetition rate difference modulation. PID, proportional-integral-derivative controller; LPF, low-pass filter; PD, photodetector; OC, optical coupler; AWG, arbitrary waveform generator; BPF, band-pass filter; SUT, sample under test; PC, polarization controller; BPD, balanced photodetector; DAQC, digital acquisition card;}
\label{fig:3}
\end{figure}

To further demonstrate the stability of our source and its potential for broadband DCS, we utilized it to measure the transmission spectrum of a high-Q microresonator. However, a fundamental trade-off exists between bandwidth and acquisition speed, according to the Nyquist sampling theorem, the following relationship exists between the maximum measurable bandwidth $\Delta v_{\text{max}}$, the repetition rate $f_{\text{rep}}$, and the repetition rate difference $\Delta f_{\text{rep}}$: 

\begin{equation}
\Delta v_{\text{max}} \leq f^2_{\text{rep}} / 2 \Delta f_{\text{rep}}.
\end{equation}

a sufficiently small repetition rate difference $\Delta f_{\text{rep}}$ is required to capture a broad spectral range without aliasing. An ultra-small $\Delta f_{\text{rep}}$ leads to a substantial reduction in the spectral acquisition rate, ultimately limiting the SNR after coherent averaging and the system's ability to capture transient processes. To overcome this challenge and fully exploit the unique advantage of the zero-crossing $\Delta f_{\text{rep}}$ tunability in our laser, we implemented a dynamic $\Delta f_{\text{rep}}$ control system. The experimental setup and its operating principle are illustrated in Fig.~\ref{fig:3}.

Figure~\ref{fig:3}(a) depicts the experimental configuration of the dynamic dual-comb system. The upper section illustrates the laser control unit, where an arbitrary waveform generator (AWG) generates square-wave control signals to modulate the pump power, thereby driving the $\Delta f_{\text{rep}}$ to switch dynamically between positive and negative values. To ensure precise monitoring and stabilization of the $\Delta f_{\text{rep}}$ state, a portion of the optical signals from both laser directions is detected by photodetectors (PD1 and PD2, Thorlabs PDA05CF2). The real-time $\Delta f_{\text{rep}}$ signal is extracted via a mixer and utilized in a feedback control loop to lock the average repetition rate difference to zero.

The core mechanism of the dynamic modulation scheme is conceptually illustrated in Fig.~\ref{fig:3}(b). To obtain stable and continuous IGMs in the time domain, the system must satisfy the prerequisite of a "zero mean repetition rate difference." This condition ensures that, over long timescales, the CW and CCW pulse trains maintain quasi-synchronization, avoiding cumulative relative walk-off. However, on shorter timescales, driven by the square-wave signal from the AWG, the $\Delta f_{\text{rep}}$ periodically switches between high (positive) and low (negative) levels. This dynamic modulation causes the two pulse trains to behave like a "pendulum," continuously approaching, overlapping (generating interference), and receding from each other in the time domain. As depicted in the temporal pulse waveforms of Fig.~\ref{fig:3}(b), each time the two pulse trains scan across one another, an IGM is generated within a period of $1 / 2f_{\text{mod}}$, which would otherwise require a much longer duration $1 /\Delta f_{\text{rep}}$ in static mode. This mechanism substantially enhances the acquisition rate of the dual-comb system.

The spectroscopic measurement segment, based on the aforementioned principle, is depicted in the lower panel of Fig.~\ref{fig:3}(a). One comb (COMB1) serves as the signal arm, passing through a bandpass filter (BPF) and the sample under test (SUT). The other comb (COMB2) acts as the local oscillator (LO) arm, with its optical path adjusted by an optical delay line (ODL) to ensure temporal overlap between the pulse trains. The BPF, centered at 1560 nm with a 10-dB bandwidth of $\sim$16 nm, is employed to suppress the steep spectral "pedestals" of the output, thereby yielding a relatively flattened spectrum. The SUT is a high-Q microresonator with GHz-level mode bandwidths, chosen to effectively characterize the measurement resolution of our system. After recombining in an optical coupler (OC), the multi-heterodyne beat signal is captured by a balanced photodetector (Thorlabs PDB425C-AC). A 27 MHz low-pass filter (LPF) is used to isolate the fundamental dual-comb beat notes, which are finally digitized and processed by a 16-bit high-precision data acquisition (DAQ) card (Alazar PCIe8916M).

\begin{figure}[t]
\centering\includegraphics[width=13.5cm]{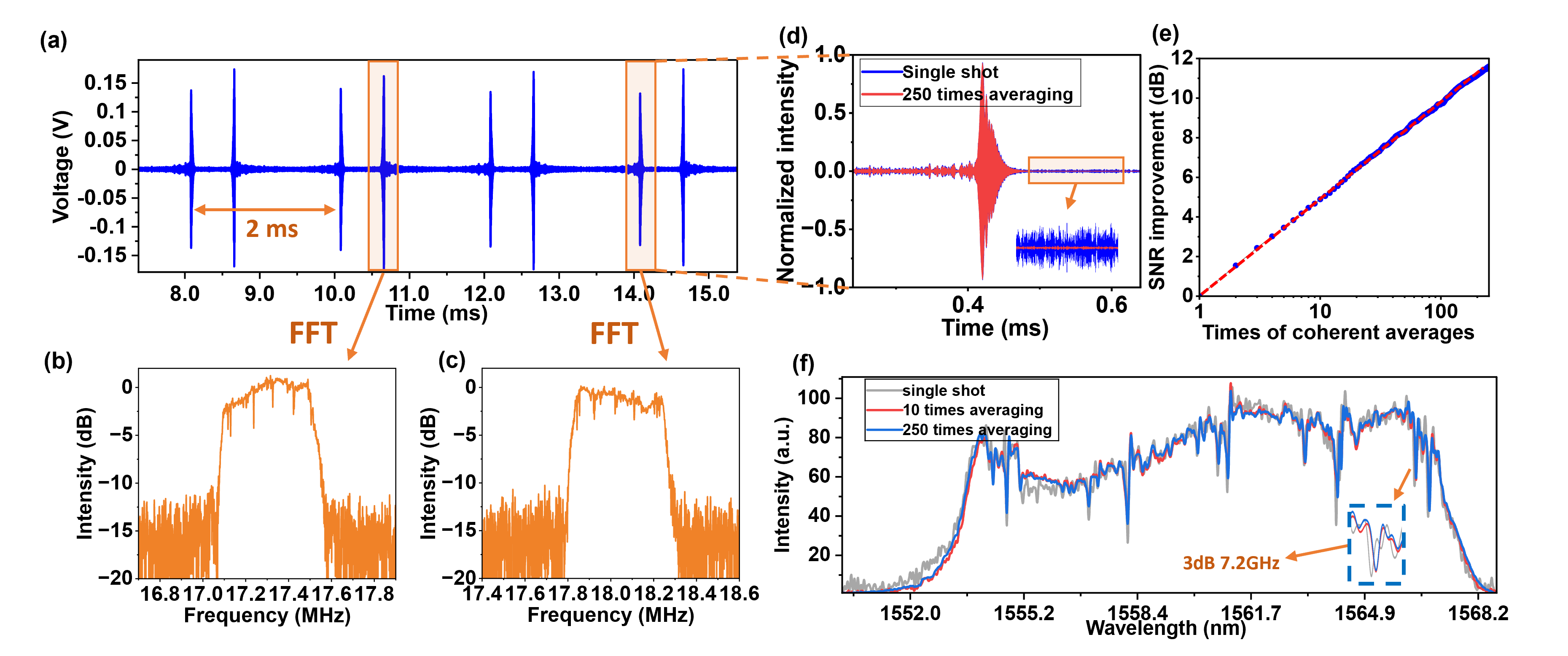}
\caption{ (a) Temporal IGMs of the DCS system under dynamic modulation. (b) and (c) Optical spectra retrieved via FFT of the IGMs generated during the forward and backward scanning processes, respectively. (d) Comparison between single-shot and coherently averaged temporal IGMs. (e) SNR as a function of the number of coherent averages; the dashed red line represents the theoretical linear trend. (f) Microresonator transmission spectra obtained with varying numbers of coherent averages.}
\label{fig:4}
\end{figure}

Figure~\ref{fig:4}(a) records the continuous-time interference waveforms output by the system at a modulation frequency of 500 Hz. Within the 2 ms modulation period, a pair of symmetrically distributed IGMs is clearly observed, corresponding to the forward and backward reciprocal scanning (walk-off) processes of the dual-comb pulses. At this stage, the $\Delta f_{\text{rep}}$ during the high and low modulation states is approximately 5 Hz, representing a nearly two-order-of-magnitude enhancement in acquisition speed. The single-shot dual-comb spectra, retrieved via Fast Fourier Transform (FFT) from these two symmetric interference frames, are presented in Figs.~\ref{fig:4}(b) and \ref{fig:4}(c). Notably, as the pump power fluctuates with the modulation signal during dynamic tuning, the instantaneous gain differs slightly between the forward and backward scans. This leads to subtle variations in the spectral envelopes of the two directions, while the underlying spectral absorption information remains consistent.

To verify the system's mutual coherence and the efficacy of our algorithm, a 0.5-second data segment containing 250 interference cycles was analyzed. A phase-correction algorithm based on the cross-ambiguity function \cite{hebert2019self} was implemented for the coherent averaging of IGMs from the same scanning direction (odd frames). Figure~\ref{fig:4}(d) compares a single-shot IGM with one obtained after 250 coherent averages. As shown in the insets, the peak intensity of the interference signal remains constant, while the background noise is significantly suppressed through averaging. The SNR analysis in Fig.~\ref{fig:4}(e) further quantifies this improvement. Within 250 averages, the SNR exhibits a near-linear scaling relationship with the square root of the average count. After 250 averages, the SNR increases by 11.59 dB (close to the theoretical limit of 12 dB) relative to the single-shot pattern. This indicates that our system maintains robust mutual coherence for over 0.5 s under dynamic $\Delta f_{\text{rep}}$ modulation. The resulting dual-comb spectra, covering a 10-dB measurement bandwidth of ~16 nm, are displayed in Fig.~\ref{fig:4}(f). The comparison highlights that while single-shot spectra are limited by SNR-induced fluctuations, the profiles after 10 and 250 averages show high consistency. To quantify the resolution, the inset of Fig.~\ref{fig:4}(f) focuses on a narrow resonance dip. After 250 averages, the measured 3-dB bandwidth is $\sim $7.2 GHz, suggesting that the system resolution within a 0.5 s acquisition time is better than 7.2 GHz. Higher-Q microresonators could potentially unveil even narrower resonances, further challenging the system's resolution limit. These results confirm that our dynamic DCS system maintains high mutual coherence and long-term stability at a 500 Hz refresh rate and a 16 nm bandwidth, and extends the actual measurement bandwidth to 5 times the theoretical non-aliasing bandwidth, providing a reliable platform for high-precision, broadband, and rapid spectroscopic measurements.

\section{Conclusion}
In summary, we have demonstrated an all-fiber, 1550 nm bidirectional single-cavity dual-comb fiber laser based on a synergistic Lyot filter and NPR mode-locking mechanism. By incorporating DCF to establish a net-normal dispersion regime, we have realized, for the first time in this band, high-energy bidirectional dissipative soliton output with pulse energies of 2.7 nJ and 1.5 nJ. This performance represents the state-of-the-art for single-cavity dual-comb systems at 1550 nm, with 10-dB spectral bandwidths exceeding 20 nm in both directions. Crucially, the $\Delta f_{\text{rep}}$ can be flexibly tuned across the zero-point by simply adjusting the pump power or intracavity polarization.

To address the inherent trade-off between spectral bandwidth and acquisition speed in DCS, we implemented a dynamic $\Delta f_{\text{rep}}$ scanning scheme via 500 Hz pump modulation. Experimental results confirm that the system maintains high-speed acquisition at 500 Hz (a nearly two-order-of-magnitude improvement over static measurements) while capturing a 16 nm spectral bandwidth. Supported by a phase-correction algorithm, the system preserves robust mutual coherence over a 0.5 s integration time, yielding a spectral resolution better than 7.2 GHz. This novel light source combines high energy, exceptional stability, and compactness. By overcoming the performance bottlenecks of conventional DCS, this work lays a solid foundation for ultra-broadband and rapid dual-comb sensing in the 1.5 $\mu$m region, with significant potential for engineered applications such as real-time multi-gas sensing and molecular fingerprinting.


\bibliography{sample}

\begin{thebibliography}{10}
\newcommand{\enquote}[1]{``#1''}

\bibitem{diddams2020optical}
S.~A. Diddams, K.~Vahala, and T.~Udem, \enquote{Optical frequency combs: Coherently uniting the electromagnetic spectrum,} {\protect\JournalTitle{Science}} \textbf{369}, eaay3676 (2020).

\bibitem{coddington2016dual}
I.~Coddington, N.~Newbury, and W.~Swann, \enquote{Dual-comb spectroscopy,} {\protect\JournalTitle{Optica}} \textbf{3}, 414--426 (2016).

\bibitem{picque2019frequency}
N.~Picqu{\'e} and T.~W. H{\"a}nsch, \enquote{Frequency comb spectroscopy,} {\protect\JournalTitle{Nature Photonics}} \textbf{13}, 146--157 (2019).

\bibitem{ideguchi2016kerr}
T.~Ideguchi, T.~Nakamura, Y.~Kobayashi, and K.~Goda, \enquote{Kerr-lens mode-locked bidirectional dual-comb ring laser for broadband dual-comb spectroscopy,} {\protect\JournalTitle{Optica}} \textbf{3}, 748--753 (2016).

\bibitem{liao2020dual}
R.~Liao, H.~Tian, W.~Liu, \emph{et~al.}, \enquote{Dual-comb generation from a single laser source: principles and spectroscopic applications towards mid-ir—a review,} {\protect\JournalTitle{Journal of Physics: Photonics}} \textbf{2}, 042006 (2020).

\bibitem{zeng2013bidirectional}
C.~Zeng, X.~Liu, and L.~Yun, \enquote{Bidirectional fiber soliton laser mode-locked by single-wall carbon nanotubes,} {\protect\JournalTitle{Optics Express}} \textbf{21}, 18937--18942 (2013).

\bibitem{olson2018bi}
J.~Olson, Y.~Ou, A.~Azarm, and K.~Kieu, \enquote{Bi-directional mode-locked thulium fiber laser as a single-cavity dual-comb source,} {\protect\JournalTitle{IEEE Photonics Technology Letters}} \textbf{30}, 1772--1775 (2018).

\bibitem{li2020bidirectional}
B.~Li, J.~Xing, D.~Kwon, \emph{et~al.}, \enquote{Bidirectional mode-locked all-normal dispersion fiber laser,} {\protect\JournalTitle{Optica}} \textbf{7}, 961--964 (2020).

\bibitem{grelu2012dissipative}
P.~Grelu and N.~Akhmediev, \enquote{Dissipative solitons for mode-locked lasers,} {\protect\JournalTitle{Nature photonics}} \textbf{6}, 84--92 (2012).

\bibitem{chong2007all}
A.~Chong, W.~H. Renninger, and F.~W. Wise, \enquote{All-normal-dispersion femtosecond fiber laser with pulse energy above 20 nj,} {\protect\JournalTitle{Optics letters}} \textbf{32}, 2408--2410 (2007).

\bibitem{camenzind2025ultra}
S.~L. Camenzind, B.~Sierro, B.~Willenberg, \emph{et~al.}, \enquote{Ultra-low noise spectral broadening of two combs in a single andi fiber,} {\protect\JournalTitle{APL Photonics}} \textbf{10} (2025).

\bibitem{rieker2014frequency}
G.~B. Rieker, F.~R. Giorgetta, W.~C. Swann, \emph{et~al.}, \enquote{Frequency-comb-based remote sensing of greenhouse gases over kilometer air paths,} {\protect\JournalTitle{Optica}} \textbf{1}, 290--298 (2014).

\bibitem{nakajima2019all}
Y.~Nakajima, Y.~Hata, and K.~Minoshima, \enquote{All-polarization-maintaining, polarization-multiplexed, dual-comb fiber laser with a nonlinear amplifying loop mirror,} {\protect\JournalTitle{Optics Express}} \textbf{27}, 14648--14656 (2019).

\bibitem{chernysheva2016isolator}
M.~Chernysheva, M.~A. Araimi, H.~Kbashi, \emph{et~al.}, \enquote{Isolator-free switchable uni-and bidirectional hybrid mode-locked erbium-doped fiber laser,} {\protect\JournalTitle{Optics express}} \textbf{24}, 15721--15729 (2016).

\bibitem{zhao2018polarization}
X.~Zhao, T.~Li, Y.~Liu, \emph{et~al.}, \enquote{Polarization-multiplexed, dual-comb all-fiber mode-locked laser,} {\protect\JournalTitle{Photonics Research}} \textbf{6}, 853--857 (2018).

\bibitem{schliesser2005frequency}
A.~Schliesser, M.~Brehm, F.~Keilmann, and D.~W.~v. der Weide, \enquote{Frequency-comb infrared spectrometer for rapid, remote chemical sensing,} {\protect\JournalTitle{Optics express}} \textbf{13}, 9029--9038 (2005).

\bibitem{kim2010high}
Y.~Kim and D.-S. Yee, \enquote{High-speed terahertz time-domain spectroscopy based on electronically controlled optical sampling,} {\protect\JournalTitle{Optics letters}} \textbf{35}, 3715--3717 (2010).

\bibitem{shi2022high}
Y.~Shi, D.~Hu, R.~Xue, \emph{et~al.}, \enquote{High speed time-of-flight displacement measurement based on dual-comb electronically controlled optical sampling,} {\protect\JournalTitle{Optics Express}} \textbf{30}, 8391--8398 (2022).

\bibitem{lyot1944filtre}
B.~Lyot, \enquote{Le filtre monochromatique polarisant et ses applications en physique solaire,} {\protect\JournalTitle{Annales d'Astrophysique, Vol. 7, p. 31}} \textbf{7}, 31 (1944).

\bibitem{han2022flexible}
D.~Han, L.~Mei, Z.~Hui, \emph{et~al.}, \enquote{Flexible wavelength-, pulse-controlled mode-locked all-fiber laser based on a fiber lyot filter,} {\protect\JournalTitle{Optics Express}} \textbf{30}, 41271--41278 (2022).

\bibitem{li2025comprehensive}
P.~Li, A.~Wang, J.~Du, \emph{et~al.}, \enquote{Comprehensive noise analysis for counter-propagating all-normal dispersion (candi) fiber laser,} {\protect\JournalTitle{APL Photonics}} \textbf{10} (2025).

\bibitem{hebert2019self}
N.~B. H{\'e}bert, V.~Michaud-Belleau, J.-D. Desch{\^e}nes, and J.~Genest, \enquote{Self-correction limits in dual-comb interferometry,} {\protect\JournalTitle{IEEE Journal of Quantum Electronics}} \textbf{55}, 1--11 (2019).

\end{thebibliography}






\end{document}